\documentclass[a4paper,11pt]{article}
\usepackage{pos}
\usepackage{graphicx,hyperref,url}
\usepackage{subfigure}
\usepackage{float}
\usepackage{tikz}
\usetikzlibrary{trees}
\usetikzlibrary{decorations.pathmorphing}
\usetikzlibrary{decorations.markings}
\usetikzlibrary{patterns}
\tikzset{
   global scale/.style={
      scale=#1,
      every node/.append style={scale=#1}},
   photon/.style={decorate, decoration={snake}, draw=red},
   nucleon/.style={draw=black, postaction={decorate},
      decoration={markings,mark=at position .55 with{\arrow[draw=black]{>}}}},
   pion/.style={draw=blue, postaction={decorate},
      decoration={markings,mark=at position .55 with{\arrow[draw=blue]{}}}},
}

\title{A dispersive analysis of low energy pion photo- and electroproduction}
\author*[a]{Xiong-Hui Cao}
\author[b]{Yao Ma}
\affiliation[a]{Department of Physics and State Key Laboratory of Nuclear Physics and Technology,\\
  Peking University,
Beijing 100871, P. R. China}

\affiliation[b]{School of Fundamental Physics and Mathematical Sciences,\\
Hangzhou Institute for Advanced Study, UCAS, Hangzhou, 310024, P. R. China}

\emailAdd{xionghuicao@pku.edu.cn}

\abstract{A dispersive representation based on unitarity and analyticity is used to study the low energy $\gamma N \to \pi N\ \text{and}\ \gamma^* N \to \pi N$ partial wave amplitudes.
Final state interactions of the $\pi N$ system are critical to this analysis.
The left-hand cut contribution is estimated by invoking baryon chiral perturbation theory results,
while the right-hand cut contribution responsible for final state interaction effects is taken into account via an Omn\`{e}s formalism with elastic phase shifts as input.
It is found that a good numerical fit can be achieved with only one subtraction parameter, and the experimental data of the multipole amplitudes $E_{0+}, S_{0+}$ in the energy region below the $\Delta (1232)$ are well described when the photon virtuality $Q^2 \leq 0.1 \mathrm{GeV}^2$.
Furthermore, we extend the partial wave amplitudes to the second Riemann sheet to extract the couplings of the subthreshold resonance $N^*(890)$.
Its couplings extracted from the multipole amplitudes $E_{0+}, S_{0+}$ are comparable to those of the $N^* (1535)$ resonance.}

\FullConference{%
  10th International workshop on Chiral Dynamics - CD2021 \\
  15-19 November 2021 \\
  Online 
}

\begin{document}
\maketitle

\section{Introduction}

An important challenge in low energy QCD is to understand the nature of nucleons and nucleon resonances.
The electromagnetic couplings of these states have long been recognized as an probe of information for understanding hadron physics.
In recent decades there have been many measurements of single pion photo- and electroproduction off nucleons, accumulating a wealth of extensive data (for a review, 
 see~\cite{Aznauryan:2011qj}). 
Based on the recent low energy experiments, partial wave (PW) analyses have been performed to study the underlying structure of the nucleon resonances, which can help us exploring the properties of QCD in the non-perturbative region.

Along with previous work in the literature~\cite{Ronchen:2014cna}, $\gamma N \to \pi N$ and $\gamma^* N \to \pi N$ processes are helpful to determine the basic properties of nucleon resonances, such as pole structures, anomalous magnetic and electromagnetic couplings.
Furthermore, these studies will serve as the basis for further photon nucleus investigation which can reveal the structure and electromagnetic properties of nucleus, for instance, the investigation of light meson photoproduction off deuteron based on scattering on nucleon~\cite{Titov:2007fc}.

At low energies, baryon chiral perturbation theory (B$\chi$PT) has become a successful tool to explore the relevant processes, with regard to the near threshold region and the low partial wave, which usually gives the precise accord~\cite{Bernard:1993bq, Hilt:2013fda, GuerreroNavarro:2019fqb, GuerreroNavarro:2020kwb}.
However, B$\chi$PT only works well near the threshold and
fails at slightly higher energies. 
So some unitarity methods are necessarily adopted in order to suppress the contributions from high energy region and recast unitarity of the amplitude.

The $N/D$ method is widely used to unitarize the photoproduction amplitude and to fit the experimental data in
couple channel $\gamma N \to \pi N$ and $\pi N\to \pi N$ reactions~\cite{Gasparyan:2010xz}.
Meanwhile, dynamical couple channel models, such as unitarity isobar model, Dubna–Mainz–Taipei (DMT) model~\cite{Kamalov:2001qg} and Jülich-Bonn-Washington (JBW) model~\cite{Mai:2021vsw}, are proposed to give a comprehensive understanding about plenty of partial waves with energy up to $2~\text{GeV}$.

Based on B$\chi$PT calculations of pion photo- and electroproduction, we have performed a dispersive analysis on these processes~\cite{Ma:2020hpe, Cao:2021kvs}.
Final state interaction is estimated by Omn\`es solution~\cite{Omnes:1958hv} in single channel approximation. 
In order to achieve such a dispersive analysis, efforts have been made in understanding the complicated analytic structure of the amplitudes.
The Omn\`es formalism is used in the study of scattering off nucleon in this talk, which have recently been successfully applied to other processes, like $\gamma\gamma \to \pi\pi$~\cite{Mao:2009cc}, $\eta\to 3\pi$~\cite{Albaladejo:2017hhj}, heavy meson decays~\cite{Yao:2018tqn}, and even the study of XYZ states~\cite{Chen:2019mgp}.

Single pion electroproduction off the nucleon is described by $e\left(l_{1}\right)+N\left(p_{1}\right) \rightarrow e\left(l_{2}\right)+N\left(p_{2}\right)+\pi(q)$.
We only consider the lowest contribution, i.e., one-photon-exchange approximation, which means the sub-process $\gamma^{*}(k)+N\left(p_{1}\right) \rightarrow N\left(p_{2}\right)+\pi(q)$ in this talk.
In this approximation, the invariant amplitude $\mathcal{M}=\epsilon_\mu \mathcal{M}^\mu$, takes the following form:
\begin{align}
    \mathcal{M}^{\mu}=-i e\left\langle N \pi\left|J^{\mu}(0)\right| N\right\rangle=\bar{u}\left(p_{2}\right)\left(\sum_{i=1}^{6} A_{i} M_{i}^{\mu}\right) u\left(p_{1}\right)\ ,\quad \epsilon_\mu=e \frac{\bar{u}\left(l_{1}\right) \gamma_{\mu} u\left(l_{2}\right)}{k^{2}}\ .
\end{align}
Six independent structures are considered here taking advantage of electromagnetic current conservation.

For PW amplitudes, the CGLN amplitudes $\mathbf{F}$ ($\mathcal{F}_i$) are conveniently used $\mathcal{M}^{\mu}=\frac{4 \pi \sqrt{s}}{m_{N}} \chi_{2}^{\dagger} \mathbf{F} \chi_{1}$.
They are defined in the center of mass (CM) frame using Coulomb gauge. The matrix element $\mathcal{F}_i$ reads~\cite{Chew:1957tf}
\begin{align}
\mathbf{F}=& i \sigma \cdot \boldsymbol{b} \mathcal{F}_{1}+\sigma \cdot \hat{\boldsymbol{q}} \sigma \cdot(\hat{\boldsymbol{k}} \times \boldsymbol{b}) \mathcal{F}_{2}+i \sigma \cdot \hat{\boldsymbol{k}} \hat{\boldsymbol{q}} \cdot \boldsymbol{b} \mathcal{F}_{3}+i \sigma \cdot \hat{\boldsymbol{q}} \hat{\boldsymbol{q}} \cdot \boldsymbol{b} \mathcal{F}_{4} \nonumber\\
&-i \sigma \cdot \hat{\boldsymbol{q}} b_{0} \mathcal{F}_{7}-i \sigma \cdot \hat{\boldsymbol{k}} b_{0} \mathcal{F}_{8}, \quad b_{\mu}=\epsilon_{\mu}-\frac{\boldsymbol{\epsilon} \cdot \hat{\boldsymbol{k}}}{|\boldsymbol{k}|} k_{\mu}\ .
\end{align}
It is convenient to work with CGLN amplitudes giving simple representations in terms of electric, magnetic and scalar (longitudinal) multipoles and derivatives of Legendre
polynomials~\cite{Berends:1967vi}
\begin{align}
    \begin{aligned}
    &\mathcal{F}_{1}=\sum_{l=0}^{\infty}\left\{\left[l M_{l+}+E_{l+}\right] P_{l+1}^{\prime}(x)+\left[(l+1) M_{l-}+E_{l-}\right] P_{l-1}^{\prime}(x)\right\}\ , \\
    &\mathcal{F}_{2}=\sum_{l=1}^{\infty}\left\{(l+1) M_{l+}+l M_{l-}\right\} P_{l}^{\prime}(x)\ , \\
    &\mathcal{F}_{3}=\sum_{l=1}^{\infty}\left\{\left[E_{l+}-M_{l+}\right] P_{l+1}^{\prime \prime}(x)+\left[E_{l-}+M_{l-}\right] P_{l-1}^{\prime \prime}(x)\right\}\ , \\
    &\mathcal{F}_{4}=\sum_{l=2}^{\infty}\left\{M_{l+}-E_{l+}-M_{l-}-E_{l-}\right\} P_{l}^{\prime \prime}(x)\ , \\
    &\mathcal{F}_{7}=\sum_{l=1}^{\infty}\left[l S_{l-}-(l+1) S_{l+}\right] P_{l}^{\prime}(x)=\frac{\left|k^{*}\right|}{k_{0}^{*}} \mathcal{F}_{6}\ , \\
    &\mathcal{F}_{8}=\sum_{l=0}^{\infty}\left[(l+1) S_{l+} P_{l+1}^{\prime}(x)-l S_{l-} P_{l-1}^{\prime}(x)\right]=\frac{\left|k^{*}\right|}{k_{0}^{*}} \mathcal{F}_{5}\ ,
\end{aligned}
\end{align}
where $x=\cos \theta=\hat{\boldsymbol{q}} \cdot \hat{\boldsymbol{k}}$.
The multipoles $E_{l \pm}, M_{l \pm}, \text{and} S_{l \pm}$ are
functions of the CM total energy $W$ and the photon virtuality
$Q^2=-q^2$, and refer to transversal electric, magnetic, and scalar (longitudinal) transitions ($L_{l \pm}=\left(k_{0} /|\boldsymbol{k}|\right) S_{l \pm}$), respectively.
Another common set of amplitudes are helicity amplitudes, linearly related to the CGLN amplitudes.
The relation between the two can be found, e.g., in Ref.~\cite{Aznauryan:2011qj}.

The isospin structure of the scattering amplitude can be
written as
\begin{align}
    \mathcal{M}\left(\gamma^{*}+N \rightarrow \pi^{a}+N\right)=\chi_{2}^{\dagger}\left\{\delta^{a 3} \mathcal{M}^{(+)}+i \epsilon^{a 3 b} \tau^{b} \mathcal{M}^{(-)}+\tau^{a} \mathcal{M}^{(0)}\right\} \chi_{1}\ ,
\end{align}
where $\tau_a$ ($a=1,2,3$) are Pauli matrices.
We can define the isospin transition amplitudes by $M^{I, I_{3}}\left(M^{\frac{3}{2}, \pm \frac{1}{2}}, M^{\frac{1}{2}, \pm \frac{1}{2}}\right)$, where $\left\{I, I_{3}\right\}$ denote isospin of the final $\pi N$ system.
The isospin transition amplitudes can be obtained from $A^{(\pm)}$
and $A^{(0)}$ via
\begin{align}
M^{\frac{3}{2}, \frac{1}{2}} &=M^{\frac{3}{2},-\frac{1}{2}}=\sqrt{\frac{2}{3}}\left(M^{(+)}-M^{(-)}\right)\ , \\
M^{\frac{1}{2}, \frac{1}{2}} &=-\sqrt{\frac{1}{3}}\left(M^{(+)}+2 M^{(-)}+3 M^{(0)}\right)\ , \\
M^{\frac{1}{2},-\frac{1}{2}} &=\sqrt{\frac{1}{3}}\left(M^{(+)}+2 M^{(-)}-3 M^{(0)}\right)\ .
\end{align}

\section{A dispersive analysis on pion electroproduction off nucleons}

We only consider the lowest PW $E_{0+},S_{0+}$ in $S_{11}$ ($L_{2I~2J}$) channel in this talk.
The treatment of the chiral interaction using SU(2) B$\chi$PT up to $\mathcal{O}(p^2)$ is described in detail in Ref.~\cite{Ma:2020hpe}.
Considering the Watson`s final state theorem and using the dispersion relation, the multipoles $E_{0+},S_{0+}$ (herewith abbreviated as $\mathcal{M}$ ) in $S_{11}$ channel are related to the $\mathcal{O}(p^2)$ chiral amplitudes $\mathcal{M}_L$~\cite{Babelon:1976kv}:
\begin{align}\label{eq1}
    \mathcal{M}(s)=\mathcal{M}_{L}(s)+\Omega(s)\left(\frac{s^n}{\pi} \int_{s_R}^{\infty} \frac{\sin\delta(s^\prime) \mathcal{M}_{L}\left(s^{\prime}\right)}{|\Omega(s^\prime)|s^{\prime n}\left(s^{\prime}-s\right)} \mathrm{d} s^{\prime}+\mathcal{P}_{n-1}(s)\right)\ ,
\end{align}
where the subscript `$L$' means it only contains left-hand cut contribution; $s_{R,L}=(m_\pi \pm m_N)^2$ is $\pi N$ threshold (pseduthreshold); $\mathcal{P}_{n}(s)$ is a $n$'s-order subtraction polynomial.
In the case of single channel problem, auxiliary function $\Omega (s)$ has a well-known analytic representation—the Omn\`es solution~\cite{Omnes:1958hv}:
\begin{align}
    \Omega(s)=\exp \left(\frac{s}{\pi} \int_{s_{R}}^{\infty} \frac{\delta\left(s^{\prime}\right)}{s^{\prime}\left(s^{\prime}-s\right)} \mathrm{d} s^{\prime}\right)\ ,
\end{align}
where $\delta(s)$ is the $\pi N\ S_{11}$ phase shift, in accordance with the Watson's final state theorem.

Applicability of the Omn\`es formalism (\ref{eq1}) relies on the ability to separate the amplitude into two pieces, the one having only left-hand cuts and the other having only a right-hand one. 
This, a priori, is not the case if the left-hand cuts overlapped with the unitary cut.
Therefore, we review our calculations and used the method in Ref.~\cite{Kennedy:1962} to analyze the singularity structure of the amplitudes.
This method relies on the feasibility of double spectral representation.
To be specific, the dynamical singularities of amplitudes can be obtained by investagating the dispersion integral.
As shown in Fig.~\ref{fig1}, all cuts are classified in the case of $Q^2=0$ as follows:
 \begin{itemize}
    \item[\uppercase\expandafter{\romannumeral 1})]  Unitarity cut, $s \in\left[s_{R}, \infty\right)$ on account of $s$-channel continuous spectrum;
    \item[\uppercase\expandafter{\romannumeral 2})]  $t$-channel cut, 1.~arc stems from $4 m_\pi^{2} \leq t \leq 4 m_N^{2}$;
    2.~$s \in(-\infty, 0]$ corresponds to $t\geq 4m_N^2$;
    \item[\uppercase\expandafter{\romannumeral 3})]  $u$-channel cut, $s \in\left(-\infty, s_u=\frac{m_N\left(m_N^2-m_\pi^2-m_\pi m_N\right)}{m_\pi+m_N}\right]$ due to $u\geq s_R$;
    \item[\uppercase\expandafter{\romannumeral 4})]  Pole, $s=m_N^2$, due to $t$-channel pion exchange and $u$-channel nucleon exchange.
\end{itemize}
Also for $Q^2 \neq 0$ case ($Q^2 \ll m_N^2$):
\begin{itemize}
    \item[\uppercase\expandafter{\romannumeral 1})]  Unitarity cut, $s \in\left[s_{R}, \infty\right)$ on account of $s$-channel continuous spectrum;
    \item[\uppercase\expandafter{\romannumeral 2})]  $t$-channel cut, 1.~arc stems from $4 m_\pi^{2} \leq t \leq 4 m_N^{2}$;
    2.~$s \in(-\infty, 0]$ corresponds to $t\geq 4m_N^2$;
    \item[\uppercase\expandafter{\romannumeral 3})]  $u$-channel cut, $s \in\left(-\infty, s_u=\frac{m_N^{3}-m_\pi^{2} m_N-m_\pi\left(m_N^{2}+Q^{2}\right)}{m_\pi+m_N}\right]$ due to $u\geq s_R$;
    \item[\uppercase\expandafter{\romannumeral 4})]  Cut due to $t$-channel pion exchange, with branch points located at $0,C_t,C_t^\dagger$;
    \item[\uppercase\expandafter{\romannumeral 5})]  Cut due to $u$-channel nucleon exchange, with branch points located at $0,C_u,C_u^\dagger$.
\end{itemize}
In the following we only give the positions and parameters in Fig.~\ref{fig1} related to the virtual photonproduction and the real photon case is obtained easily by setting $Q^2=0$.
The positions of the two pairs of branch point and the parameters of the arc are written as
\begin{align}
    &C_{t}, C_{t}^\dagger=M^{2}-\frac{Q^{2}}{2}\pm i \sqrt{4 M^{2} Q^{2}-m^{2} Q^{2}-\frac{Q^{4}}{4}+\frac{M^{2}}{m^{2}} Q^{4}}\ , \\
    &C_{u},C_{u}^\dagger=M^{2}-\frac{1}{2} \frac{m^{2}}{M^{2}} Q^{2}\pm i \sqrt{4 m^{2} Q^{2}-\frac{m^{2}}{M^{2}} m^{2} Q^{2}+\frac{m^{2}}{M^{2}} Q^{4}-\frac{1}{4}\left(\frac{m^{2}}{M^{2}}\right)^{2} Q^{4}}\ , \\
    &R^2=(m_N^2+Q^2)(m_N^2-m_\pi^2)\ , \\
    &s_c=m_N^2-\frac{Q^2}{2}-\frac{3 m_\pi^2}{2}\ .
\end{align}
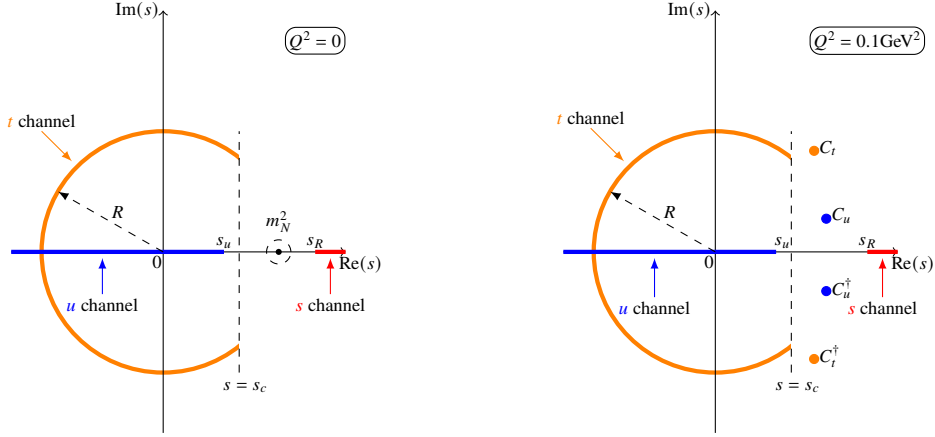
\begin{figure}
    \subfigure{
        \begin{minipage}[b]{0.985\linewidth}
    
        \begin{minipage}[b]{0.48\linewidth}\centering
    \begin{tikzpicture}[global scale=0.4]
    
        \node[draw, rounded corners] at(5,7)  {\LARGE$Q^2=0$};

        \draw[ultra thick,orange](0,0) circle (4);
        \fill [white] (2.5,-4)to(4.5,-4)to(4.5,4)to(2.5,4)to(2.5,-4);
        \draw[-latex,orange] (-4,4) -- (-3,3); 
        \node[above] at (-4,4){\LARGE\({\color{orange} t}\ \text{channel}\)};    
        \draw[dashed](2.5,-4)to(2.5,4);
        \node[above] at (2.7,-4.8){\LARGE\(s=s_c\)};       
        
        \draw[->](-5,0)to(6,0);
        \draw[->](0,-6)to(0,8);
        
        \draw[ultra thick, blue](2,0)to (-5,0);
        \node[above] at (2,0){\LARGE\(s_u\)};
        \draw[-latex,blue] (-2,-1.4) -- (-2,-0.2); 
        \node[below] at (-2,-1.4){\LARGE\({\color{blue} u}\ \text{channel}\)};  

        \draw[ultra thick, red](5,0)to (6,0);
        \node[above] at (5,0){\LARGE\(s_R\)};
        \draw[-latex,red] (5.5,-1.4) -- (5.5,-0.2); 
        \node[below] at (5.5,-1.4){\LARGE\({\color{red} s}\ \text{channel}\)};  

        \fill[black](3.8,0) circle (0.1);
        \node[above] at (3.8,0.5){\LARGE\(m_N^2\)};
        \draw[dashed] (3.8,0) circle (0.4);
        
        \node[below] at (-0.2,0){\LARGE\(0\)};
        
        \draw[dashed,-latex] (0,0) -- (-3.464,2); 
        \node[above] at (-1.5,1){\LARGE\(R\)};
        \node[left] at (0,8){\LARGE $\mathrm{Im}(s)$};
        \node[below] at (6.5,0){\LARGE $\mathrm{Re}(s)$};
    \end{tikzpicture}
        \vspace{0.02cm}
        \hspace{0.02cm}
        \end{minipage}
        \begin{minipage}[b]{0.48\linewidth}\centering
    \begin{tikzpicture}[global scale=0.4]

        \node[draw, rounded corners] at(5,7)  {\LARGE$Q^2=0.1\text{GeV}^2$};

        \draw[ultra thick,orange](0,0) circle (4);
        \fill [white] (2.5,-4)to(4.5,-4)to(4.5,4)to(2.5,4)to(2.5,-4);
        \draw[-latex,orange] (-4.1,4.1) -- (-3,3); 
        \node[above] at (-4.1,4.1){\LARGE\({\color{orange} t}\ \text{channel}\)};   
        \draw[dashed](2.5,-4)to(2.5,4);
        \node[above] at (2.7,-4.8){\LARGE\(s=s_c\)};       
        
        \draw[->](-5,0)to(6,0);
        \draw[->](0,-6)to(0,8);  
        \node[above,right] at (3.1, 3.45){$\bullet$\LARGE\(C_t\)};
        \filldraw[orange] (3.25, 3.35) circle (.15);
        \node[above,right] at (3.1, -3.45){$\bullet$\LARGE\(C_t^\dagger\)};   
        \filldraw[orange] (3.25, -3.55) circle (.15);   
        
        \draw[ultra thick, blue](2,0)to (-5,0);
        \node[above] at (2.2,0){\LARGE\(s_u\)};
        \draw[-latex,blue] (-2,-1.5) -- (-2,-0.2); 
        \node[below] at (-2,-1.5){\LARGE\({\color{blue} u}\ \text{channel}\)};  
        
        \node[above,right] at (3.5, 1.2){$\bullet$\LARGE\(C_u\)};  
        \filldraw[blue] (3.65, 1.1) circle (.15);   
        \node[above,right] at (3.5, -1.2){$\bullet$\LARGE\(C_u^\dagger\)};      
        \filldraw[blue] (3.65, -1.3) circle (.15);   
        
        \draw[ultra thick, red](5,0)to (6,0);
        \node[above] at (5,0){\LARGE\(s_R\)};
        \draw[-latex,red] (5.5,-1.5) -- (5.5,-0.2); 
        \node[below] at (5.5,-1.5){\LARGE\({\color{red} s}\ \text{channel}\)};  
        
        \node[below] at (-0.2,0){\LARGE\(0\)};
        
        \draw[dashed,-latex] (0,0) -- (-3.464,2); 
        \node[above] at (-1.5,1){\LARGE\(R\)};
        \node[left] at (0,8){\LARGE $\mathrm{Im}(s)$};
        \node[below] at (6.5,0){\LARGE $\mathrm{Re}(s)$};
    \end{tikzpicture}
        \vspace{0.02cm}
        \hspace{0.02cm}
        \end{minipage}
                       
        \end{minipage}

            }
    \caption{Singularities in the complex $s$ plane. 
    The left panel is real photon case, and the right  is the case of virtual photon.
    }\label{fig1}
\end{figure}

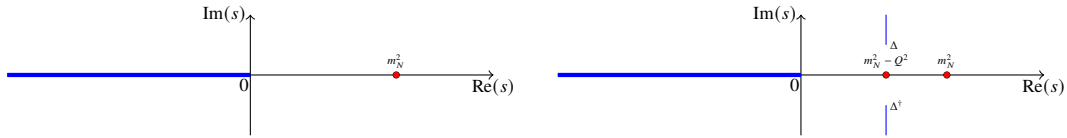
\begin{figure}[H]
    \subfigure{
        \begin{minipage}[b]{0.985\linewidth}
    
        \begin{minipage}[b]{0.48\linewidth}\centering
    \begin{tikzpicture}[global scale = 0.4]
            \draw[->](-8,0)to(8,0);
            \draw[->](0,-2)to(0,2);
    
            \draw[ultra thick, blue](-8,0)to (0,0);        
            
            \draw (4.8,0) circle (0.1);
            \fill[red](4.8,0) circle (0.1);
            \node[above] at (4.8,0.1){\(m_N^2\)};
            
            \node[below] at (-0.2,0){\LARGE\(0\)};
            \node[left] at (0,2){\LARGE $\mathrm{Im}(s)$};
            \node[below] at (8,0){\LARGE $\mathrm{Re}(s)$};
        \end{tikzpicture}
        \vspace{0.02cm}
        \hspace{0.02cm}
        \end{minipage}
        \begin{minipage}[b]{0.48\linewidth}\centering
    \begin{tikzpicture}[global scale = 0.4]
        \draw[->](-8,0)to(8,0);
        \draw[->](0,-2)to(0,2);

        \draw[ultra thick, blue](-8,0)to (0,0);        
        
        \draw (4.8,0) circle (0.1);
        \fill[red](4.8,0) circle (0.1);
        \node[above] at (4.8,0.1){\(m_N^2\)};
        \draw (2.8,0) circle (0.1);
        \fill[red](2.8,0) circle (0.1);
        \node[above] at (2.8,0.1){\(m_N^2-Q^2\)};
        
        \draw[blue](2.8,1)to (2.8,2);
        \draw[blue](2.8,-1)to (2.8,-2);
        \node[right] at (2.8,1){\(\Delta\)};
        \node[right] at (2.8,-1){\(\Delta^\dagger\)};
        
        \node[below] at (-0.2,0){\LARGE\(0\)};
        \node[left] at (0,2){\LARGE $\mathrm{Im}(s)$};
        \node[below] at (8,0){\LARGE $\mathrm{Re}(s)$};
    \end{tikzpicture}
        \vspace{0.02cm}
        \hspace{0.02cm}
        \end{minipage}
                       
        \end{minipage}

            }
    \caption{The kinematical singularities in the $s$ plane.
    The left panel describes the real photonproduction, and the right  is the case of virtual photon.
    The pole $s=m_N^2-Q^2$ is derived from the requirement of gauge invariance and $\Delta=m_N^2-Q^2+i 2MQ$.}\label{fig2}
\end{figure}

Compared to real photon case, the singularities of virtual photon case include some additional branch cuts in the complex $s$ plane, rather than on the real axis.
A more detailed study of real and virtual photoproduction is discussed in Refs.~\cite{Ma:2020hpe, Cao:2021kvs}, where it is pointed out that there are additional kinematical singularities from relativistic kinematics, especially in an inelastic scattering process.
The possible singularities in our PW analysis ($S_{11}$ channel) are displayed in Fig.~\ref{fig2}.
It can be concluded that all cuts do not cover the unitarity cut~\cite{Cao:2021kvs}, so the whole singularities can be divided into right-hand cuts and left-hand cuts.

\begin{figure}
    \centering 
    \includegraphics[width=11cm]{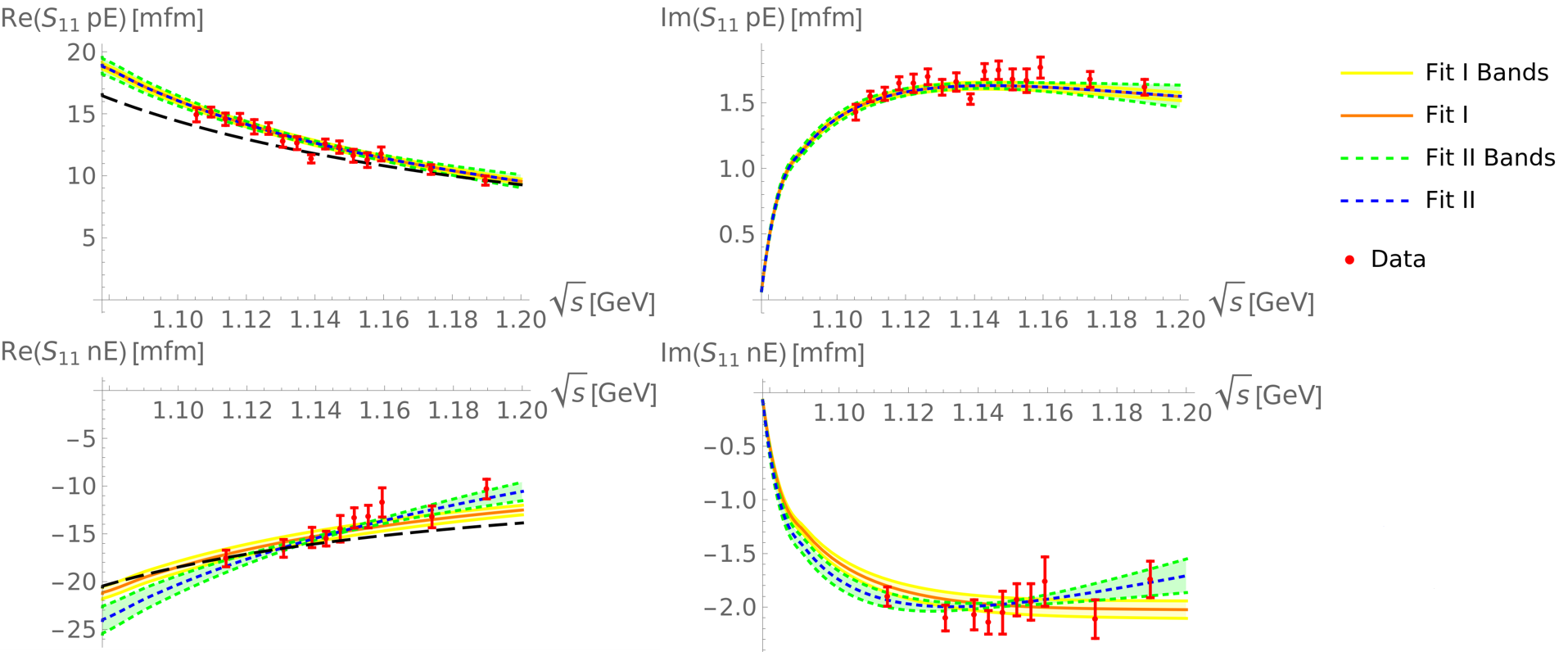}
    \caption{Real and imaginary parts of the $E_{0+}$ multipole within $p$ and $n$ targets.
    Fit I is the case of once subtraction, and fit II is the case of twice subtraction.}\label{fig3}
\end{figure}

By virtue of Eq.~(\ref{eq1}), the $s$-wave multipoles ($S_{11}$ channel) are obtained by fitting $E_{0+} (S_{0+})$ of proton ($p$) and neutron ($n$) targets simultaneously to data, derived from SAID PW analysis group~\cite{SAID} for photoproduction and DMT2001 and MAID2007 for electroproduction.
The phase shifts are extracted from the $\pi N$ $S$-matrix given in the Roy-Steiner analysis~\cite{Hoferichter:2015hva}.
The fit results of photoproduction for both targets are displayed in Fig.~\ref{fig3}.
It is found that a once subtraction $\mathcal{P}_0=a$ is enough.
Since we notice that the effects of twice subtractions ($\mathcal{P}_1=as+b$) does not get a improved physical outputs and the fit parameters $a$ and $b$ of fit II are highly negative correlated, with a correlation coefficient that is nearly $-1$. 
Thus, fit I is more advisable.

\begin{figure}
    \centering
    \includegraphics[width=16cm]{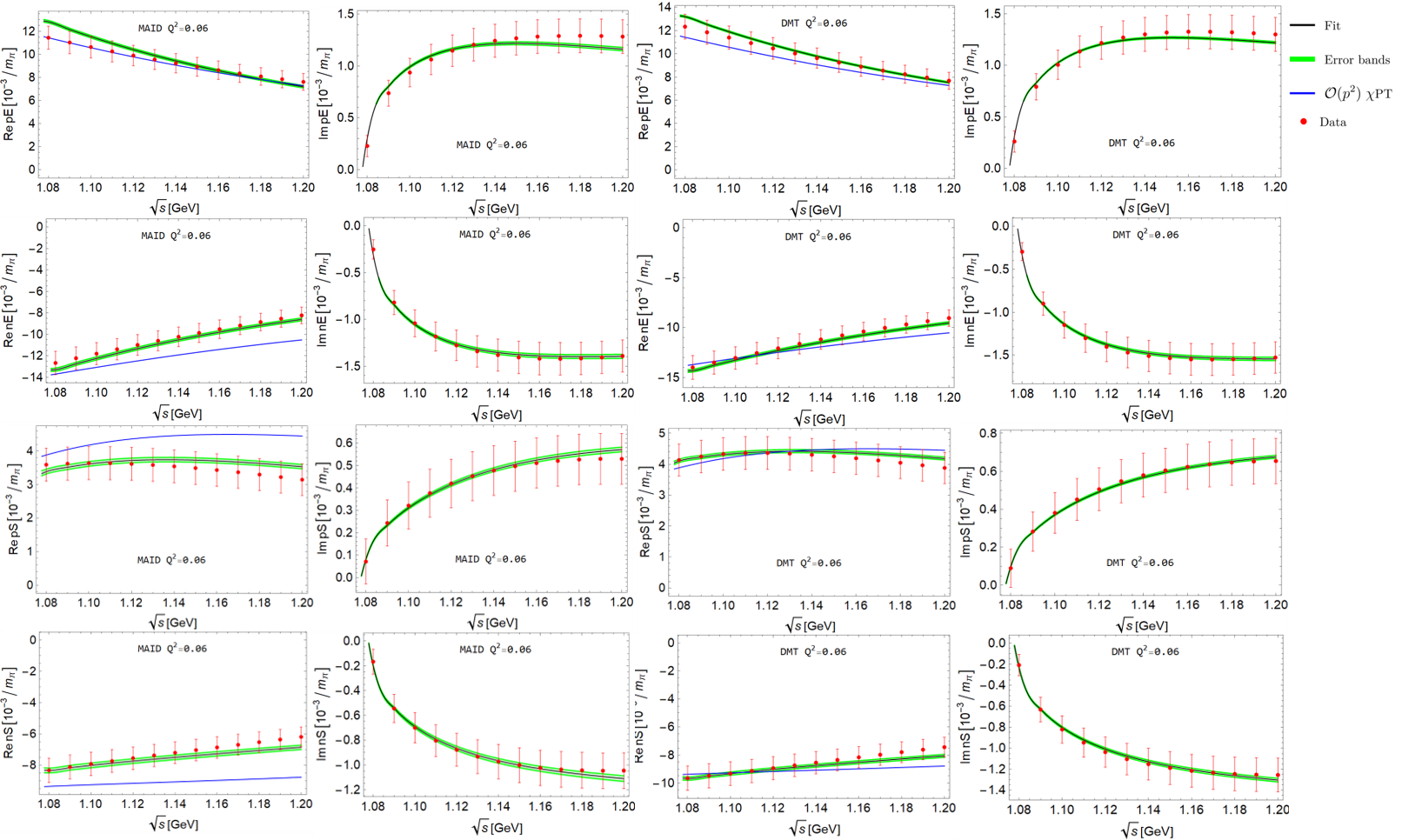}            
    \caption{$E_{0+}$ and $S_{0+}$ for $p$ and $n$ targets: the ``data” are from MAID (two left columns) and DMT (two right columns), respectively.
    }\label{fig4}
\end{figure}

As for electroproduction, the fit procedure is similar as before.
The only subtle thing worth noting is the fact that the subtraction constant $a(Q^2)$ is $Q^2$ dependent in principle.
However, we further assume $a(Q^2)$ to be independent of $Q^2$, since it mainly depends on the couplings between photon and light vector mesons such as $\rho$ and $\omega$ due to vector meson dominance.
Similar discussions on this issue in the mesonic sector ($\gamma\gamma \to \pi\pi$) can be found in Ref.~\cite{Moussallam:2013una}.
The fit results with $Q^2=0.06\mathrm{GeV}^2$ are plotted in Fig.~\ref{fig4} both for MAID and DMT models.

It is convincing that no matter what data are used, the fit results of multipoles are similar. 
The results can illustrate that Omn\`es-like method is very powerful and effective in low energy regions and low $Q^2$ regions.
The method is superior to pure $\mathcal{O}(p^2)$ perturbation calculations in the sense that the dispersive representation can generate the corresponding
imaginary parts. 
Further, it is hard to compare the $\mathcal{O}(p^4)$ $\chi$PT results~\cite{Hilt:2013fda, GuerreroNavarro:2019fqb, GuerreroNavarro:2020kwb} and our calculations. 
In our calculation, we only use the left-hand part contribution extracted from the $\mathcal{O}(p^2)$ amplitude. 
In principle, an $\mathcal{O}(p^4)$ calculation is
advantageous compared with an $\mathcal{O}(p^2)$ calculation. 
But in a perturbation calculation, unitarization effects are not taken into account, which are automatically fulfilled in
our scheme.

\section{Electromagnetic couplings of the subthreshold resonance}

In Refs.~\cite{Wang:2017agd, Wang:2018gul, Wang:2018nwi}, evidences are found on the possible existence of a subthresthod resonance named $N^*(890)$ in the $S_{11}$ channel using the method proposed in Ref.~\cite{Xiao:2000kx, He:2002ut, Zheng:2003rw, Zhou:2004ms, Zhou:2006wm} (for a review, see Ref.~\cite{Yao:2020bxx}).
Further, some follow-up work confirmed this discovery in different ways such as $K$-matrix~\cite{Ma:2020sym} and $N/D$ method~\cite{Li:2021tnt}.
It could be surprising to notice that the pole position, is located below the threshold $s_R$.
However, a subthreshold pole on the second Riemann (RS) sheet with such a large imaginary part is allowed and its existence does not violate the principle of QFT such as causality, etc..
By analyzing the whole pion production process, the above estimates of amplitudes can be utilized to obtain the electromagnetic couplings of $N^*(890)$.

The PW electroproduction amplitude on the second RS can be deduced via
\begin{align}\label{eq2}
    \mathcal{M}_l^{\mathrm{II}}(s)=\frac{\mathcal{M}_l(s)}{\mathcal{S}_l(s)}\ ,
\end{align}
where $\mathcal{S}_l(s)$ corresponds to the $S$ matrix of $\pi N$ scattering, with the same quantum numbers as $\mathcal{M}_l(s)$.
In the vicinity of a second RS pole $s_R$, $\mathcal{S}(s)$ can be approximated by $\mathcal{S}(s) \simeq \mathcal{S}^{\prime}\left(s_{\mathrm{R}}\right)\left(s-s_{\mathrm{R}}\right)$.
For such a pole, we can define the residues as
\begin{align}
    \mathcal{M}^{\mathrm{II}}\left(s \rightarrow s_{\mathrm{R}}\right)\simeq\frac{g_{\gamma N} g_{\pi N}}{s-s_{\mathrm{R}}}\ ,
\end{align}
$g_{\gamma N}$ and $g_{\pi N}$ denote the $\gamma N$ and $\pi N$ couplings, respectively.
Compared with Eq.~(\ref{eq2}), we obtain
\begin{align}\label{eq3}
    g_{\gamma N} g_{\pi N} \simeq \frac{\mathcal{M}_{l}\left(s_{\mathrm{p}}\right)}{\mathcal{S}_{l}^{\prime}\left(s_{\mathrm{p}}\right)}\ ,
\end{align}
The $\pi N$ coupling can also be extracted from $g_{\pi N}^{2} \simeq \mathcal{T}_{l}\left(s_{\mathrm{p}}\right) / \mathcal{S}_{l}^{\prime}\left(s_{\mathrm{p}}\right)$, where $\mathcal{T}_l(s)$ is the corresponding PW $\pi N$ amplitude.
Using the above formulae, we can calculate the (virtual-)photon decay amplitudes $A_{1/2}^{\text {pole}}$ ($S_{1/2}^{\text {pole}}$) at the pole position of $N^*$~\cite{Tiator:2016btt}:
\begin{align}\label{eq4}
    A_{1 / 2}^{\text {pole }}\left(Q^{2}\right)=g_{\gamma N}^{E} \sqrt{\frac{3 \pi W_{\mathrm{p}}}{m_{N} k_{\mathrm{p}}^{\mathrm{cm}}}}, \quad S_{1 / 2}^{\text {pole }}\left(Q^{2}\right)=g_{\gamma N}^{S} \sqrt{\frac{3 \pi W_{\mathrm{p}}}{2 m_{N} k_{\mathrm{p}}^{\mathrm{cm}}}}\ ,
\end{align}
where $k^{\text{cm}}=(W^2-m_N^2)/(2W)$ is the real photon equivalent energy in CM frame.

\begin{figure}[H]
    \centering
    \subfigure
    {
    \begin{minipage}[b]{0.985\linewidth}
    
    \begin{minipage}[b]{0.45\linewidth}
    \includegraphics[width=6.5cm]{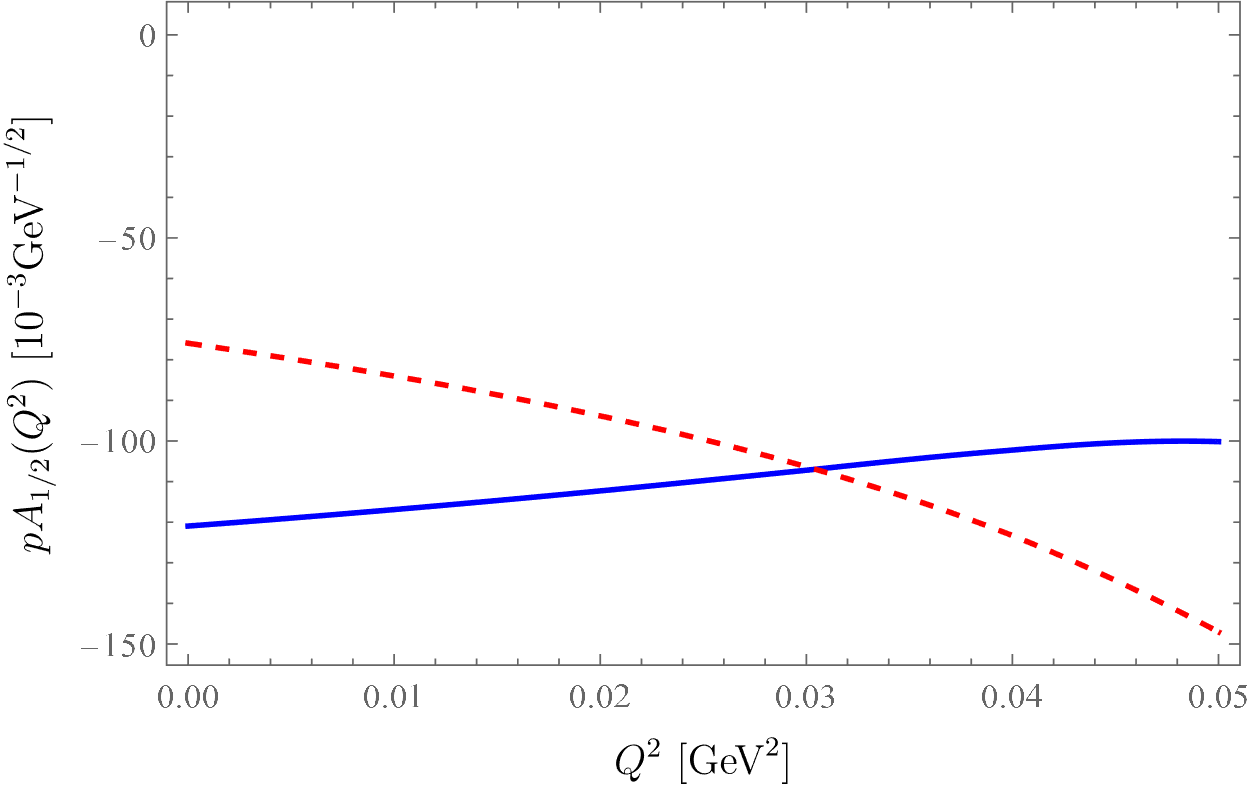}
    \vspace{0.02cm}
    \hspace{0.02cm}
    \end{minipage}
    \begin{minipage}[b]{0.45\linewidth}
    \includegraphics[width=6.5cm]{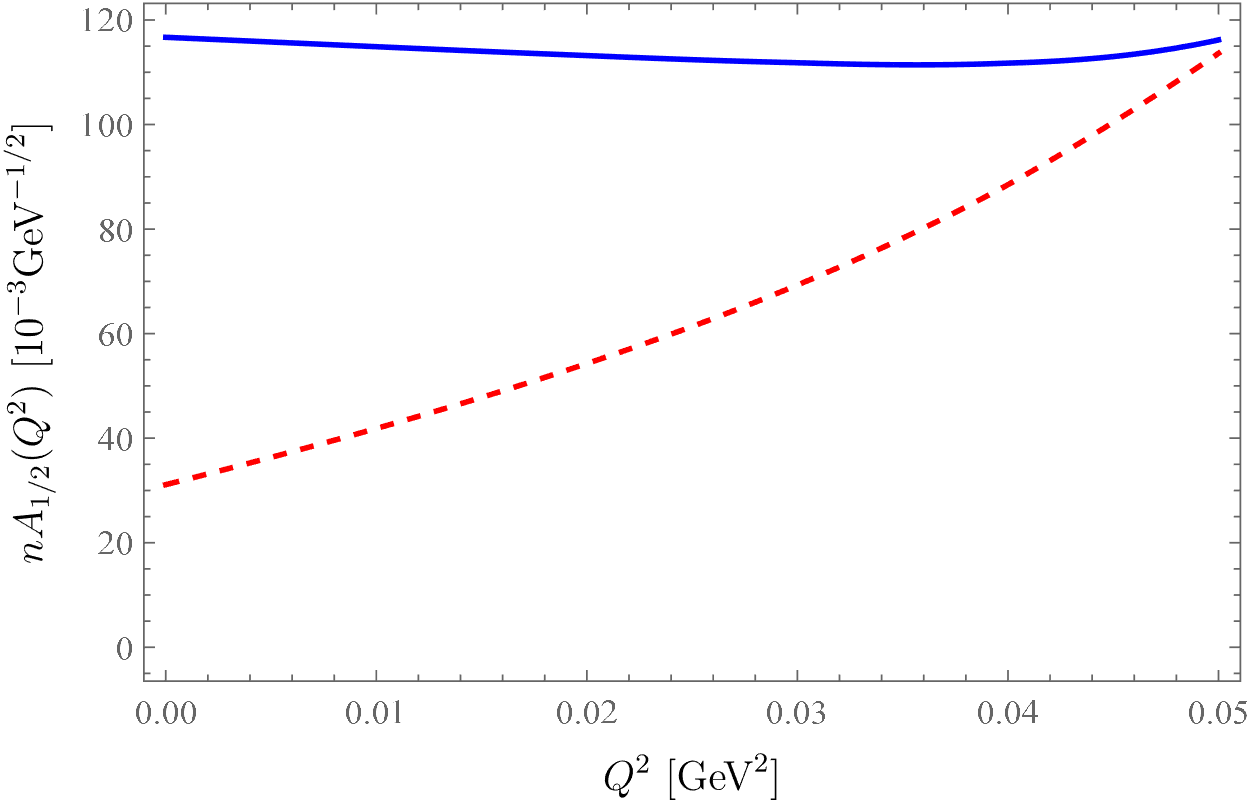}
    \vspace{0.02cm}
    \hspace{0.02cm}
    \end{minipage}

    \begin{minipage}[b]{0.45\linewidth}
    \includegraphics[width=6.5cm]{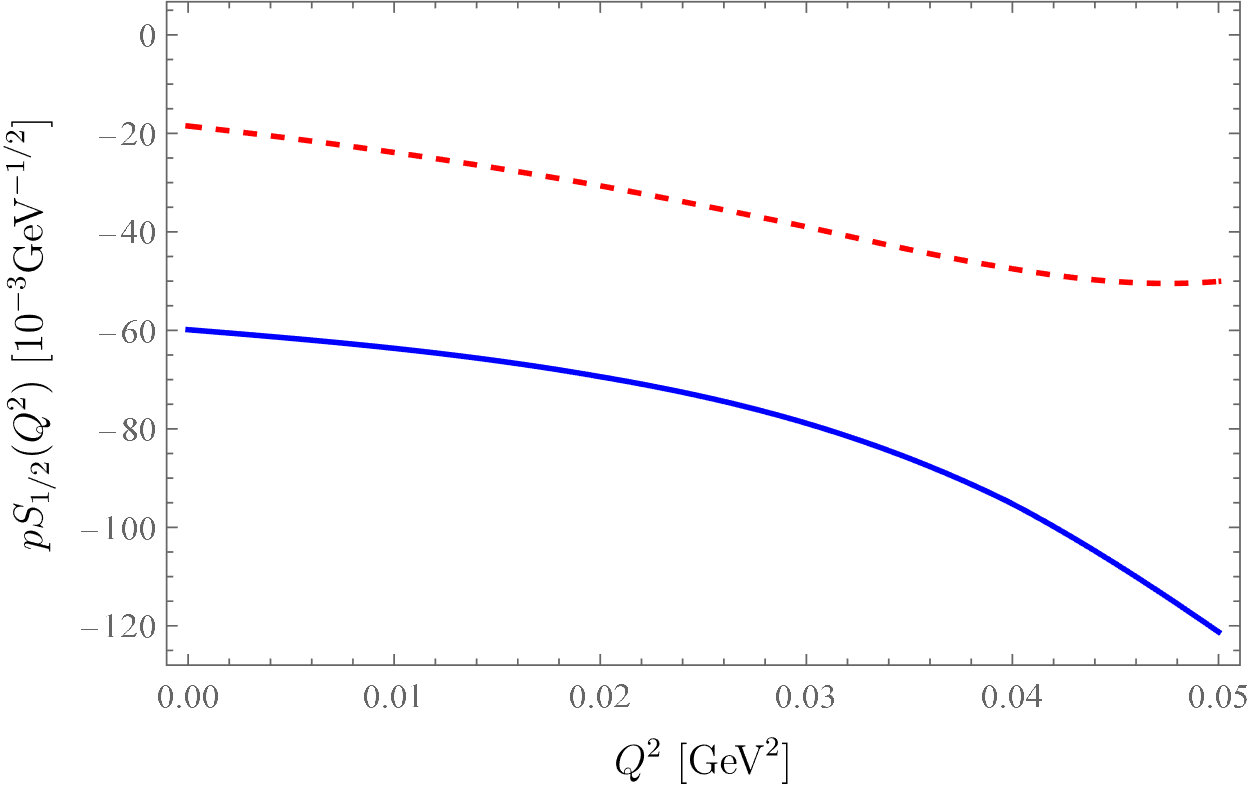}
    \vspace{0.02cm}
    \hspace{0.02cm}
    \end{minipage}
    \begin{minipage}[b]{0.45\linewidth}
    \includegraphics[width=6.5cm]{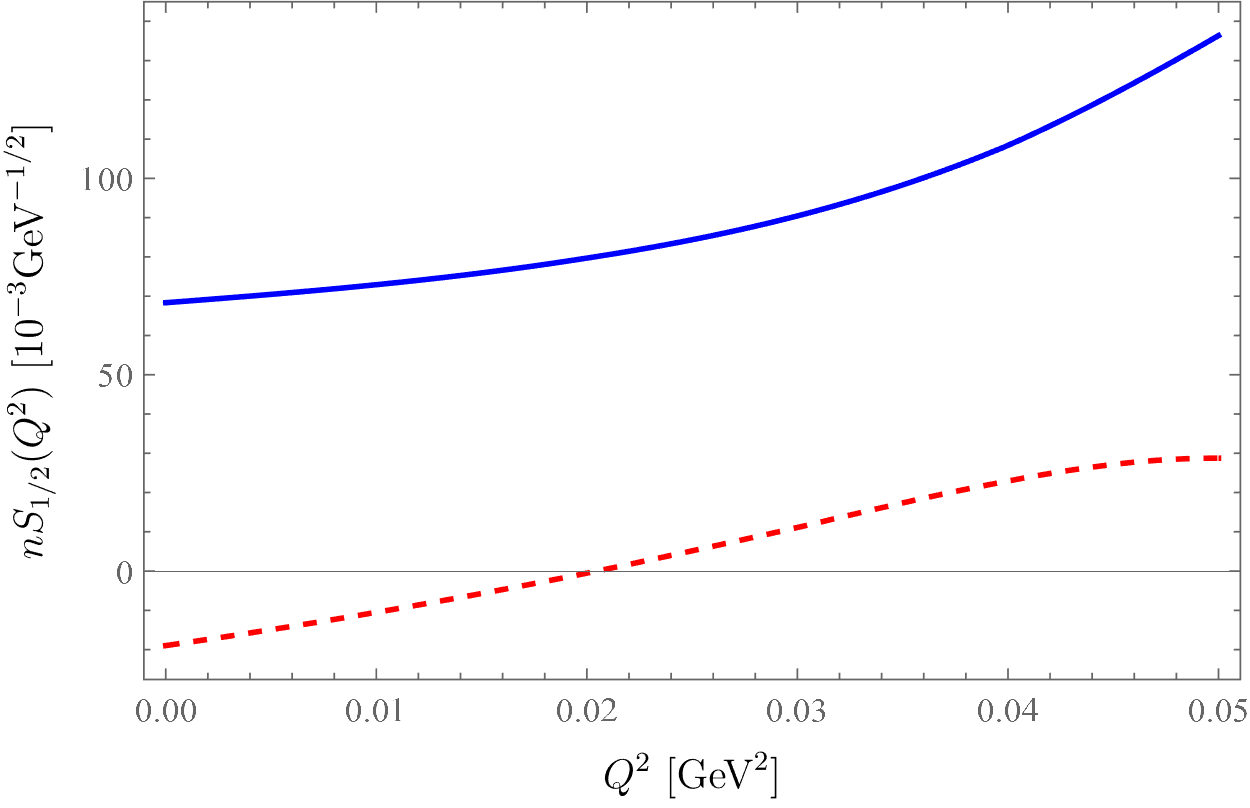}
    \vspace{0.02cm}
    \hspace{0.02cm}
    \end{minipage}
                       
    \end{minipage}

    }
    \caption{The blue solid line and the red dashed line represent real and imaginary parts of virtual photon decay amplitudes at pole position $\sqrt{s}=0.882-0.190 i$, respectively.}\label{fig5}
\end{figure}

According to Eq.~(\ref{eq3}), the residues of $N^{*}(890)$, $g_{\gamma N} g_{\pi N}$, can be extracted from multipoles $E_{0+}, S_{0+} .$ 
In the meantime, $g_{\pi N}^{2}$ can be computed by using $g_{\pi N}^{2} \simeq \mathcal{T}_{l}\left(s_{\mathrm{p}}\right) / \mathcal{S}_{l}^{\prime}\left(s_{\mathrm{p}}\right)$, which was already obtained in Ref.~\cite{Wang:2017agd}. 
We employed our fit of MAID data, and chose $\sqrt{s}=0.882-0.190 i$ for pole position to extract pole residues. 
$\mathcal{T}\left(s_{\mathrm{p}}\right)$ can be obtained through $\mathcal{S}\left(s_{\mathrm{p}}\right)=1+2 i \rho_{\pi N}\left(s_{\mathrm{p}}\right) \mathcal{T}\left(s_{\mathrm{p}}\right)=0$, where $\rho_{\pi N}=\sqrt{(s-s_R)(s-s_L)}/s$.
The values of the residues and the decay amplitudes at the $N^*(890)$ and $N^*(1535)$ pole positions are presented in Tab.~\ref{tab1}.
In the case of elecproduction, the common $Q^2$ dependence virtual decay amplitudes $A_{1/2}(Q^2)$ and $S_{1/2}(Q^2)$ are also obtained in Fig.~~\ref{fig5}. 
\begin{table}
        \centering
        \footnotesize
        \begin{tabular}{lll}
        \hline \hline  & $N^*(890)$ & $N^*(1535)$ \\
         \hline $|g_{\pi N}|^2$ &  $0.2\mathrm{GeV}^2$~\cite{Wang:2017agd} &  $0.08\mathrm{GeV}^2$~\cite{Arndt:2006bf} \\
         $|g_{\gamma N}|$ &  $0.032\mathrm{GeV}$~\cite{Ma:2020hpe} &  $0.024\mathrm{GeV}$~\cite{Svarc:2014sqa} \\
         $|pA_{1/2}|$ &  $0.17\mathrm{GeV}^{-1/2}$~\cite{Ma:2020hpe} &  $0.007\mathrm{GeV}^{-1/2}$~\cite{Svarc:2014sqa} \\
        \hline \hline
        \end{tabular}
        \caption{Results of the couplings and the modulus of p target decay amplitudes.}\label{tab1}
\end{table}

In Tab.~\ref{tab1}, we just note that the magnitude of the coupling of $N^*(890)$ $g_{\pi N}$ is larger than that of $N^*(1535)$.
One can immediately obtain the residue of photoproduction using Eq.~(\ref{eq3}).
It is clear that the magnitudes of $g_{\gamma N}$ are nearly the same.
However, the magnitude of the real photon decay amplitudes $p A_{1/2}$ of $N^*(890)$ is also larger than that of $N^*(1535)$ due to the kinematical factor in Eq.~(\ref{eq4}).
Similarly, it is also the reason why the values of virtual decay amplitudes in Fig.~~\ref{fig5} are so large.
But there is a potential difficulty to be concerned.
The values of the virtual decay amplitudes are highly sensitive to the pole position of $N^*(890)$, which is derived from fitting data of $\pi N$ $S_{11}$ phase shift.
Bearing these considerations in mind, Fig.~\ref{fig5} should be treated as a qualitative order of magnitude estimation rather than a quantitative calculation.
An improved result relies on the precise location of $N^*(890)$, in prospect of accurate calculations in the future, such as Roy-Steiner analysis like $\sigma$~\cite{Caprini:2005zr} and $\kappa$~\cite{Descotes-Genon:2006sdr}.
\\
\\
The authors would like to thank Han-Qing Zheng for a close collaboration.
This work is supported in part by National Nature Science Foundations of China (NSFC) under Contracts No. 11975028 and
No. 10925522.

\end{document}